\documentclass{article}
\usepackage{dcase2022,amsmath, amssymb,graphicx,url,times,booktabs, tabularx,siunitx,paralist,multirow,balance}
\usepackage[breaklinks=true]{hyperref}
\usepackage{cleveref}[2012/02/15]
\usepackage[nolist]{acronym}
\usepackage{cite}
\usepackage[acronym]{glossaries}
\glsnogroupskiptrue
\glsdisablehyper
\makeglossaries

\newacronym{DNN}{DNN}{deep neural network}
\newacronym{FLOPs}{FLOPs}{floating point operations per second}
\newacronym{ASR}{ASR}{automatic speech recognition}
\newacronym{LLMs}{LLMs}{large-language models}
\newacronym{SED}{SED}{sound event detection}
\newacronym{SSL}{SSL}{self-supervised learning}
\newacronym{MSE}{MSE}{mean square error}
\newacronym{BCE}{BCE}{binary cross entropy}

\crefformat{footnote}{#2\footnotemark[#1]#3}

\title{DCASE 2024 Task 4:\\Sound Event Detection with Heterogeneous Data and Missing Labels}

\name{Samuele Cornell$^{1,*}$, Janek Ebbers$^{2,*}$, Constance Douwes$^{3}$,}
\secondlinename{	  
       Irene Mart\'{i}n-Morat\'{o}$^{4}$,
       Manu Harju$^{4}$,
       Annamaria Mesaros$^{4}$, Romain Serizel$^{3}$
       }
 \address{    
         $^1$Carnegie Mellon University, USA
         \quad $^2$Mitsubishi Electric Research Laboratories, USA \\
         $^3$ Universite de Lorraine, CNRS, Inria, Loria, Nancy, France
         \quad $^4$  Tampere University, Finland 
         }

\begin{document}

\ninept
\maketitle

\def\thefootnote{*}\footnotetext{These authors contributed equally to this work}\def\thefootnote{\arabic{footnote}}

\begin{sloppy}
\begin{abstract}
The Detection and Classification of Acoustic Scenes and Events Challenge Task 4 aims to advance \gls{SED} systems in domestic environments by leveraging training data with different supervision uncertainty. 
Participants are challenged in exploring how to best use training data from different domains and with varying annotation granularity (strong/weak temporal resolution, soft/hard labels), to obtain a robust SED system that can generalize across different scenarios. Crucially, annotation across available training datasets can be inconsistent and hence sound labels of one dataset may be present but not annotated in the other one and vice-versa. As such, systems will have to cope with potentially missing target labels during training.  
Moreover, as an additional novelty, systems will also be evaluated on labels with different granularity in order to assess their robustness for different applications.
To lower the entry barrier for participants, we developed an updated baseline system with several caveats to address these aforementioned problems.
Results with our baseline system indicate that this research direction is promising and is possible to obtain a stronger \gls{SED} system by using diverse domain training data with missing labels compared to training a \gls{SED} system for each domain separately.

%The systems need to be able to correctly detect the sound events present in a recorded audio clip, as well as localize the events in time. This year's task is a follow-up of DCASE 2021 Task 4, with some important novelties. The goal of this paper is to describe and motivate these new additions, and report an analysis of their impact on the baseline system. 
%We introduced three main novelties: the use of external datasets, including recently released strongly annotated clips from Audioset, the possibility of leveraging pre-trained models, and a new energy consumption metric to raise awareness about the ecological impact of training sound events detectors.  
%The results on the baseline system show that leveraging open-source pre-trained on AudioSet improves the results significantly in terms of event classification but not in terms of event segmentation.
\end{abstract}

\begin{keywords}
Sound event detection, missing labels, efficiency, weak supervision, heterogeneous data
\end{keywords}

\section{Introduction}
\label{sec:intro}

It can be argued that, with current deep learning based techniques, the ability to leverage as much training data as possible is as important as the pursue of novel (in the methodological sense) techniques.
For example, the effectiveness of modern \gls{LLMs} relies mostly on the scale of the training data rather than on their \gls{DNN} architecture. 
The same is true for \gls{ASR} models, with recent works~\cite{radford2023robust, peng2023reproducing, chen2022wavlm} demonstrating that a great deal of robustness, as well as zero-shot and emerging capabilities~\cite{radford2023robust}, come both from the scale of the model and, crucially, the size of the training set. 

However, leveraging data at scale has its own set of challenges. This is particularly true for \gls{SED} where readily available data and metadata is not readily obtainable from web sources. 
While \gls{SSL} techniques~\cite{gong2022ssast, baade2022mae, chen2023beats, huang2022masked} can help to circumvent this issue, still, supervised data is necessary for fine-tuning. 
For this latter, the only viable option right now is hand annotation, which is very expensive and difficult to scale as \gls{SED} requires temporal endpoints together with the class label. 
Hence, to lower the annotation burden, temporally \emph{weak} annotations (i.e. presence or not of a sound event inside a particular audio clip of several seconds without precise endpoints) are often used in conjunction with a smaller portion of temporally precise (i.e. \emph{strong}) annotated recordings~\cite{serizel2018large, serizel2018_DCASE}. 
These latter are particular important, as it has been demonstrated~\cite{hershey2021benefit, ronchini2021impact} that increasing the amount of strong-labeled examples brings considerably benefits in terms of performance, despite the obvious drawbacks of increasing the annotations costs.  
As such, in the recently proposed MAESTRO~\cite{martin2023training} dataset, a sliding window approach to the annotation procedure was developed. 
This approach, together with crowdsourcing, allows for better scaling in the annotation stage. In fact, in MAESTRO temporally strong labels are obtained by overlap-add of several temporally weak annotations. 

This discrepancy in the annotation temporal granularity, since 2018, has been explored extensively in the past DCASE Task 4 challenges~\cite{serizel2018_DCASE, turpault2020training, Turpault2019_DCASE, serizel2020sound, dcase2021Task4, ronchini2022description}, with DESED~\cite{serizel:hal-02355573, turpault:hal-02160855} being the main dataset used through these all past editions.

However, another crucial issue is that, between different datasets, not only the temporal granularity (temporally strong vs. weak labels) can vary but also the consistency in the annotation procedure. I.e. which classes are considered as events of interest and which are instead disregarded, or again, if annotation confidence (i.e. the use of \emph{soft} labels) is available or not.
This direction has been largely underexplored in previous DCASE Task 4 challenges but is essential towards the goal of leveraging as much as training data as possible and is the main novelty introduced this year.

\section{Motivation}
\label{sec:motivation}

%This year  aims to advance \gls{SED} research by proposing new research questions which are grounded in practical problems for the field. 

This year the DCASE Challenge Task 4 aims at addressing $2$ different aspects that are related to the aforementioned problem of leveraging diverse training data with missing and (temporally and/or posterior-wise) weak annotation.
Each of these aspects answer fundamental research questions which are formulated in the following. 

\subsection{Can we combine datasets from diverse domains with different annotations to improve performance ?}
\label{ssec:combine_datasets}

As said in Section~\ref{sec:intro}, one of the challenges of combining different datasets for \gls{SED} is the fact that the two datasets may not have consistent annotation with one another. In extreme cases, the datasets might not even share any common sound event classes.
Instead of training a \gls{SED} model on each dataset separately an intriguing approach is to just train one model on all available datasets. 
Intuitively, if the two datasets have sound classes that overlap or, at least, some classes that could be mapped from one another (e.g. when one event is a sub-class of another event~\cite{shrivastava2020mt, wichern2010ontological, shah2023approach}), then we expect that using all datasets should afford better performance compared to training a model for each separately.  
However, since annotation can be inconsistent and some events that are annotated in one dataset may be present but not annotated in the other, the training procedure and possibly even the \gls{SED} model must be modified to account for this issue. 
In Section~\ref{sec:baseline_changes} we describe how we addressed this when developing this year baseline system and in Section~\ref{ssec:missing_labels_results} we present also some results which indicate that this research direction is promising and indeed leads to large performance gains. 
Our hope is that participants will devise novel and more effective ways to address this important problem. 

\subsection{What is the best way to exploit soft labels ? Are they useful to improve performance ?}
\label{ssec:soft_labels_useful}

Some datasets, such as MAESTRO, due to their data annotation protocol, have soft labels expressing the annotators overall confidence of the presence or not of a particular sound event.
In ~\cite{martin2023training} it was shown that it is possible to train an effective \gls{SED} system using such soft labeled annotation and two possible loss functions: \gls{BCE} and \gls{MSE} were explored as well as different post-processing techniques. 
In particular, the choice of the loss function was found to affect the model performance on more rare occurring sound event classes. 
Several research questions however arise when soft labels are combined with strong labels from other datasets and with soft labels from pseudo labels obtained from the model (e.g. via mean-teacher~\cite{meanteacher}). I.e. it would be interesting to assess if annotation confidence metadata is useful for training a robust \gls{SED} system when training data is scaled and if also other approaches e.g. filtering may be helpful or not.

%Right now we do not combine so this question will be unsolved I guess....

\section{Challenge Datasets}
\label{sec:dataset}

This year challenge keeps using the DESED dataset, in order to be comparable with previous editions, but also adds MAESTRO as another dataset participants can use and on which performance will be evaluated. 
Both are described in detail in the following. 

\textbf{DESED} consists of $10$ seconds length audio clips either recorded in a domestic environment or synthesized to reproduce such an environment. It features annotated sound events from $10$ different classes: alarm\_bell\_ringing
        blender, cat, dishes, dog,
        electric\_shaver\_toothbrush, frying, running\_water, speech,
        vacuum\_cleaner. 
The synthetic part of the dataset is generated with Scaper~\cite{salamon2017scaper} with foreground events obtained from the Freesound datasets~\cite{fonseca2017freesound} while backgrounds are extracted from YouTube videos under Creative Commons license and from the Freesound subset of the MUSAN dataset~\cite{snyder2015musan}. Such synthetic set is divided into an evaluation and training part. More information is available in~\cite{serizel2020sound}. 
The real-world recording part is instead derived from AudioSet~\cite{gemmeke2017audio} and it comprises of a temporally-weakly annotated set ($1578$ clips), a totally unlabeled set ($14412$ clips) and also a strongly annotated portion from~\cite{hershey2021benefit} ($3470$ clips).  

%\footnote{For a detailed description of the DESED dataset and how it is generated the reader is referred to the original DESED article~\cite{turpault:hal-02160855} and DCASE 2021 task 4 webpage: \url{http://dcase.community/challenge2021}}.   

\textbf{MAESTRO Real}, which has been proposed in~\cite{martin2023training} and used in the past DCASE 2023 Task 4 (track B) challenge, consists of a development (6426 clips) and an evaluation part of long-form real-world recordings. This dataset contains multiple temporally-strong annotated events with soft labels from $17$ classes. However, in this challenge, out of these, only $11$ are considered in evaluation as the other $6$ do not occur with confidence over $0.5$.
These classes are: birds\_singing,
car,
people\_talking,
footsteps,
children\_voices,
wind\_blowing,
brakes\_squeaking,
large\_vehicle,
cutlery\_and\_dishes,
metro\_approaching,
metro\_leaving.
As said, this data was annotated using crowdsourcing and the procedure introduced in~\cite{Martin2023strong}, where temporally-weak labeling is used in conjunction to a sliding window approach to derive events temporal localization. Multiple annotators outputs are aggregated via MACE~\cite{hovy2013learning}. 
The recordings are derived from TUT Acoustic Scenes 2016~\cite{mesaros2016tut} dataset and are between 3 to 5 minutes long.

\section{Rules}
\label{sec:rules}
Rules are largely similar to previous year edition. However this year we allow participants to always use external data and pre-trained models~\footnote{Allowed data and model resources are listed in the challenge website} without reporting also results for approaches that are fully trained only with DESED and MAESTRO. 
Another important difference is that, this year, since we have two scenarios, we prohibit domain identification. In fact we want participants to focus on approaches that can generalize across various scenarios without a-priori knowledge of which subset of sound classes can be present.

%Participants are allowed to submit up to 4 different systems .
%Participants have to submit at least one system without ensembling.
%Participants have to submit (post-processed and unprocessed) output scores from three independent model trainings with different initialization to be able to evaluate the model performance's standard deviation.
%Participants are allowed to use external data for system development.
%Data from other task is considered external data.
%Embeddings extracted from models pre-trained on external data is considered as external data
%Another example of external data is other materials related to the video such as the rest of audio from where the 10-sec clip was extracted, the video frames and metadata.
%Datasets and models can be added to the list upon request until May 1st (as long as the corresponding resources are publicly available).
%The external dataset used during training should be listed in the YAML file describing the submission.
%Manipulation of provided training data is allowed.
%Participants are not allowed to use the public evaluation dataset and synthetic evaluation dataset (or part of them) to train their systems or tune hyper-parameters.
%Domain identification is prohibited: participant are not allowed to leverage domain information in inference whether the audio comes from MAESTRO or DESED

\section{Evaluation}

\textbf{SED evaluation} assesses a system's capability of recognizing and temporally localizing sound events.
Currently three different event-matching approaches exist namely collar-~\cite{mesaros2016metrics}, intersection-~\cite{bilen2020framework,ferroni2021improving} and segment-based~\cite{mesaros2016metrics}, which differ in the way they compare predicted and ground truth temporal locations of sound events.
In recent years, intersection-based evaluation has gained popularity as an event-based metric favoring detection of reasonably connected events, while being less sensitive to annotation ambiguities compared to collar-based evaluations.
Further, there is a high variation in SED application requirements, with some applications requiring a high recall, others a high precision, and yet others may even let the user control sensitivity.
Hence, an SED evaluation metric ideally aggregates performance over various operating modes.

Therefore, the polyphonic sound detection score (PSDS)~\cite{bilen2020framework,ebbers2022threshold} has been used as primary metric in this task since 2021.
It evaluates the normalized partial area under the PSD-ROC curve, where the PSD-ROC is the average of class-wise intersection-based ROC curves plus a penalty on inter-class standard deviation.
PSDS parameters are the detection tolerance criterion $\rho_\text{DTC}$ (the required intersection of a detected event with ground truth events to not be counted false positive (FP)), the ground truth intersection criterion $\rho_\text{GTC}$ (the required intersection of a ground truth event with non-FP detected event), the penalty weight $\alpha_\text{ST}$ on inter-class standard deviation, and the maximum FP-rate $e_\text{max}$ up to which the area under curve is computed~\footnote{Cross-trigger parameters are not mentioned as not considered this year.}.
In previous editions PSDS1 and PSDS2 have been evaluated, which differ in their parameters.
This year we are considering only PSDS1 for evaluation with $\rho_\text{DTC}=\rho_\text{GTC}=0.7$, ${\alpha_\text{ST}=1.}$, ${e_\text{max}=\SI{100}{FPs/hour}}$, as PSDS2 is tuned more as an audio tagging than an SED metric.
Event's onset and offset times required for PSDS computation, however, are only available for DESED data and classes which is why PSDS1 is only evaluated on this fraction of the evaluation set.

For MAESTRO, segment-based labels (segment length of one second) are provided, and we use the segment-based mean (macro-averaged) partial area under ROC curve (segMPAUC) as the primary metric instead with a maximum FP-rate of 0.1.
segMPAUC is computed w.r.t. hard labels (using a binarization threshold of 0.5) for the 11 classes listed in Sec.~\ref{sec:dataset}.

To have a common processing of DESED and MAESTRO data during inference, we split MAESTRO recordings, which comprise several minutes, into clips of 10 seconds with a clip overlap of 50\%.
DESED and MAESTRO clips are anonymized and shuffled in the evaluation set to prevent manual domain identification (cf. task rules in Sec.~\ref{sec:rules}).
At evaluation time, we reconstruct recording-level predictions from the MAESTRO clips by computing, for each class, a scalar posterior score in each segment.
To do so, submitted (short-time) class posterior scores are averaged over the duration of a segment and further averaging segment-level scores of the same segment from overlapping clips.

\begin{table}[tb]
\centering
\footnotesize
\begin{tabular}{c|cc}
Data \raisebox{.24ex}{\textbackslash} \raisebox{.48ex}{Classes}& DESED & MAESTRO \\
\hline
DESED   & PSDS1    & not evaluated      \\
MAESTRO     & not evaluated       & segMPAUC          \\
\bottomrule
\end{tabular}
\vspace{-2mm}
\caption{Primary metrics for different data and class fractions.}\label{tab:eval}
\vspace{-5mm}
\end{table}
Table~\ref{tab:eval} summarizes which primary metric is used on which data and class fractions.
Additionally, we report segMPAUC on DESED, collar-based F$_1$-score on DESED, and macro-averaged segment-based F$_1$-scores and error rates (ERs) for a detection threshold of $0.5$ and with optimal detection thresholds.
All metrics are evaluated using sed\_scores\_eval\footnote{\url{https://github.com/fgnt/sed_scores_eval}}.
As in previous editions, we use both the predictions from three independent training runs and bootstrapped evaluation~\cite{efron1994introduction} to compute a mean and standard deviation for each of the metrics.
Here, 20 different bootstrap samples (whereby we ensure that each clip is overall sampled equally often) are evaluated for each of the three runs yielding 60 results to compute statistics from.
As ranking metric the sum of the primary metrics' means $\overline{\text{PSDS1}}+\overline{\text{segMPAUC}}$ is used.
Note, that both metrics are taken from the same system, as, in contrast to previous editions, both metrics focus on SED here.

\textbf{Energy efficiency} is another important factor in SED systems.
Hence, as in the previous two editions, we ask participants to report the energy consumption of their system during both training and testing stages using the CodeCarbon package \cite{codecarbon}. We also ask participants to report the energy consumption for training the baseline model on $10$ epochs as well as for inference with the baseline model on the development set. This procedure has to be performed on the same hardware as used for their system training/inference such that energy consumption can be normalized among different hardware and provide fairer comparisons~\cite{ronchini2022description}. In addition, this year we ask not only CodeCarbon's total energy consumption, which is calculated as the sum of the three components (GPU, CPU, RAM), but also the energy from the GPU component alone. In fact, we found that CPU and RAM consumption were both included by CodeCarbon in previous DCASE Task 4 challenges, and are now also interested an accurate picture of the GPU energy. Having a more precise energy consumption estimation could allow to better assess the relationship between the number of multiply-accumulate (MAC) operations, the number of parameters and energy consumption from the GPU.
In Section~\ref{sec:results}, Table~\ref{tab:energy} energy consumption figures are for the baseline are reported. 

\section{DCASE 2024 Challenge Task 4 Baseline System} 
\label{sec:baseline_changes}

The baseline system is directly inherited from the previous DCASE Task 4 challenges~\cite{dcase2023Task4a,ronchini2022description} and consists of a convolutional recurrent neural network (CRNN) network
which also employs self-supervisedly learned features from BEATs pre-trained model~\cite{chen2023beats}. 
The CRNN model has a convolutional neural network (CNN) encoder of $7$ convolutional layers with batch normalization, gated linear unit and dropout, followed by a bi-directional gated recurrent unit (biGRU) layer. Before this latter, BEATs features are concatenated with the CNN extracted ones. Average pooling is applied to BEATs features prior concatenation to make the sequence length the same as the one from the CNN encoder. 
Clip-wise and frame-wise posteriors are then derived using an attention pooling~\cite{jiakai2018mean}. The 
CNN encoder is fed log-Mel filterbank energies extracted with a 128\,ms window and 16\,ms stride from 16\,kHz audio.
During training the BEATs model is kept frozen, Mixup~\cite{zhang2017mixup} regularization strategy is employed and the mean-teacher framework~\cite{meanteacher} is used in order to leverage unlabeled and weakly-labeled data. 
Baseline code and pre-trained checkpoints are available online\footnote{\href{https://github.com/DCASE-REPO/DESED_task/blob/master/recipes/dcase2024_task4_baseline/confs/pretrained.yaml}{github.com/DCASE-REPO/DESED\_task/recipes/dcase2024\_task4\_baseline}}.

%This year's challenge baseline is based on a  but includes the main novelty of having the possibility of using features extracted from pre-trained models. CRNN was already found to be the architecture, mainly taken from \cite{jiakai2018mean}, is composed of a CNN module  followed by a 2-layers bi-directional gated recurrent unit (biGRU). The CNN has 7 layers, each composed of batch normalization, gated linear unit and dropout.
%Input features are Log-Mel Filterbank Energies extracted with a 128 ms window and 16 ms stride. 
%The model is trained with the mean-teacher strategy \cite{meanteacher, jiakai2018mean} on audio data resampled at 16\,kHz and outputs one frame-wise prediction each 64 ms. 
%To leverage more effectively weakly and unlabeled labeled data, attention pooling is employed, as outlined in \cite{jiakai2018mean}, to derive clip-level predictions from frame-level predictions.
%From 2020, small changes, such as MixUp \cite{zhang2017mixup}.
% put on footnote the code 

For this year challenge we introduced two incremental improvements and, to deal with the aforementioned missing labels problem also some ad-hoc modifications to the training procedure. %In fact, DESED and MAESTRO may have event classes which belongs to one dataset and which are present in the other one but were not considered for annotation. 
Regarding the minor improvements for this year baseline we use SpecAugment-style~\cite{park2019specaugment} time-wise masking on the pre-trained model extracted features and, independently, on the features extracted from the CNN encoder. We denote this strategy as \emph{dropstep} in Section~\ref{sse:baseline_imp_res}.
Another difference is that for post-processing we employ a multi-class median filter where each class has a different median filter length. 

\subsection{Dealing with partially annotated data}

The training procedure had to be modified in several places in order to deal with the missing labels problem. 

\vspace{-0.2cm}
\subsubsection{Cross mapping sound event classes}
First, as a pre-processing step, we map some DESED events to similar classes in MAESTRO. More in detail, we have in DESED ``speech'' which is a super-class for ``people\_talking, children\_voices, announcement'' in MAESTRO, ``dishes'' which corresponds to ``cutlery\_and\_dishes'' and also ``dog'' which is a super-class for ``dog\_bark''.
Note that these mapping are from MAESTRO to DESED but not vice-versa as DESED ones are mostly super-classes of MAESTRO ones. 
Intuitively, with this strategy, when computing the loss on MAESTRO e.g. for a clip with the event ``people\_talking'' having confidence $0.5$, we also drive the network output posterior corresponding to ``speech'' class to $0.5$.
\vspace{-0.2cm}
\subsubsection{Loss computation}

The model is trained using \gls{BCE} loss function on DESED real-world strongly, synthetic and weakly labeled examples as well as on MAESTRO soft labeled examples.
\Gls{MSE} is instead used for the mean-teacher pseudo-labeling loss component which is applied on both weak and unlabeled data from DESED.
When computing the loss for both components on a particular clip we avoid computing the loss for the network outputs corresponding to the classes that do not correspond to the clip original dataset. For example, for MAESTRO, we do not compute the loss for DESED output logits except for class that have been cross-mapped as explained before. 

\subsubsection{Attention-pooling masking}
The attention pooling mechanism~\cite{jiakai2018mean} employed in the final layer of the baseline model applies the softmax function over classes. 
Before taking the softmax, the values corresponding to unlabeled classes (not belonging to the current clip dataset) are masked to minus infinite in order to prevent to attend to them. 
\vspace{-0.2cm}
\subsubsection{Mixup}

Mixup~\cite{zhang2017mixup} regularization strategy is applied for MAESTRO and DESED independently as labels are missing and the two cannot be mixed together in a reliable manner. 
\vspace{-0.2cm}
\subsection{Hyperparameters tuning}

We adopt a dual-phase approach to hyperparameters tuning in order to ease the computational burden of the overall tuning procedure. 
In the first step, we tune the network and training parameters~\footnote{Script available at: \href{https://github.com/DCASE-REPO/DESED\_task/blob/master/recipes/dcase2024\_task4\_baseline/optuna_pretrained.py}{dcase2024\_task4\_baseline/optuna\_pretrained.py}}. This requires training the model from scratch for each set of selected hyperparameters. In detail we tune the number of biGRU layers and its hidden state size, learning rate, dropout and dropstep parameters, warmup epochs and gradient clipping value. 
In a second step, the network is kept frozen and we use the best model as found in the first step and tune only the multi-class median filter. This second step requires only to perform inference on the dev-test portions of the data\footnote{The optimized class-wise median filters lengths are in \href{https://github.com/DCASE-REPO/DESED_task/blob/master/recipes/dcase2024_task4_baseline/confs/pretrained.yaml}{dcase2024\_task4\_baseline/confs/default.yaml}}. 
Such dual-phase approach allows to dramatically reduce the required number of training runs compared to tuning everything together from scratch, since a slight change in the median filter length for a particular class has a significant effect on the performance of the overall system, leading to a very noisy hyperparameter tuning procedure.
The hyperparameter tuning procedure was performed using the Optuna toolkit~\cite{akiba2019optuna} using  multi-objective tree-structured Parzen estimator~\cite{ozaki2022multiobjective} with dev-test $\text{PSDS1}+\text{segMPAUC}$ as the objective function.

% put on footnote code to hyperparameters especially for the median filter 

% describe two step hyper-params tuning 
% describe drop-step strategy. run ablation with it. 

% ablation: compute difference of activation on super class speech and sub-classes people talking and children voices. 
% also count activations of non-annotated events for each dataset
% compare activations of desed only model, maestro only model and both model.
% label density ? should we run experimnents on this ? 
%We also experimented with SpecAugment~\cite{park2019specaugment} but, in our case, it did not improve the results. 

\section{Experimental Results}\label{sec:results}
\subsection{Baseline improvements}\label{sse:baseline_imp_res}

In Table~\ref{tab:baseline_tuning} top-panel, we report an ablation study to motivate the baseline system changes described in Section~\ref{sec:baseline_changes}. 
We can observe that all the proposed changes bring substantial improvement. In particular, the dual-phase Optuna-based hyperparameter tuning (- HypTune ablations) appears to be quite effective. 
Adding a median filter (- HypTune 2 ablation, unprocessed scores) seems crucial, while having a multi-class median filter (- MC-Median ablation), improves performance only marginally. 
Compared to this latter, the dropstep regularization strategy has a more significant effect (- dropstep ablation).

\begin{table}[htb]
\centering
\footnotesize
\setlength{\tabcolsep}{6pt}
 \begin{tabular}{l|c|c}
   Model & \multicolumn{1}{c|}{PSDS1 $\uparrow$} & \multicolumn{1}{c}{segMPAUC $\uparrow$} \\
     \toprule
     & Dev-test (DESED) & Dev-test (MAESTRO)  \\
     %./exp/2024_baseline/version_19
    \hline
    Random Init & $0.0$ & $0.02$ \\
    \hline
    Baseline & $0.491$ & $0.731$ \\
   \hline
  - dropstep  &  $0.479$ & $0.706$ \\ %  version 21 
  - HypTune1  & $0.458$  & $0.669$ \\ 
  - HypTune2  & $0.391$ & $0.702$ \\ % version 22
  - MC-Median & $0.485$ & $0.714$  \\ % version 23 
  \hline
  \hline
  - DESED  & $0.0$ & $0.642 $ \\ % only maestro ./exp/2024_baseline/version_18
  - MAESTRO  & $0.483$ & $0.115$ \\ % only desed ./exp/2024_baseline/version_14
  - CrossMap  & $0.469$ & $0.722$ \\  % effect of removing the mapping %./exp/2024_baseline/version_11
 \bottomrule
 \end{tabular}
 \vspace{-2mm}
 \caption{Baseline improvements ablation study on dev-test and effect of training the system only on DESED or MAESTRO data. For MAESTRO, we used 90\% overlap when reconstructing the long-form audio.}
\vspace{-1mm}
 \label{tab:baseline_tuning}
\end{table}

\begin{table}[h!]
\centering
\footnotesize
\begin{tabular}{c|c|c|c}
\multicolumn{2}{c|}{Total kWh} & \multicolumn{2}{c}{GPU kWh} \\
\hline
Train   & Dev-test    & Train   & Dev-test    \\
\hline
1.180     & 0.119       & 0.113   & 0.013        \\
\bottomrule
\end{tabular}
 \vspace{-2mm}
\caption{Baseline energy consumption in kWh for training and inferring on the development set on one A100 (40GB)}\label{tab:energy}
 \vspace{-3mm}
\end{table}

% put table with hyperpar value 

\subsection{Leveraging heterogeneous datasets with missing labels}\label{ssec:missing_labels_results}

In Table~\ref{tab:baseline_tuning} bottom-panel we report an ablation study to assess how removing one of the two datasets (MAESTRO or DESED) affects the overall performance of the \gls{SED} system. 
We can see that, in both instances where the other dataset is removed, whether it is DESED (- MAESTRO ablation) or MAESTRO (- DESED ablation), the performance on the remaining dataset also drops. However, the performance drop is small if MAESTRO is removed. This is likely due to the fact that DESED is much larger and thus the effect of removing/adding MAESTRO is modest. 
The strategy described in Section~\ref{sec:baseline_changes} of mapping some MAESTRO classes to some DESED classes is considerably effective (- CrossMap ablation) in particular for DESED as one would expect (some MAESTRO classes are mapped to corresponding DESED super-classes).
What is rather surprising is, instead, the fact that if DESED is removed (- DESED ablation), the performance on MAESTRO drops quite dramatically. 
In fact, as described in Section~\ref{sec:baseline_changes}, during training, when both datasets are used, the loss on the classes that do not belong to the dataset from which the input audio is taken are masked, thus e.g. MAESTRO outputs are completely ignored when the input audio comes from DESED (we do not map any class from DESED to MAESTRO). 
We hypothesize that the addition of DESED data boosts significantly the performance on MAESTRO because it may help the model to learn how to extract a more meaningful and generalizable representation especially in the earlier layers of the network, acting as a regularization strategy (especially important as MAESTRO is small compared to DESED). This hypothesis may also explain why if we remove the class mapping (- CrossMap ablation) the performance on MAESTRO is still superior to using MAESTRO alone.

\section{Conclusions}

In this paper we presented the DCASE 2024 Task 4 challenge which addresses the important problem of leveraging multiple data sources for training \gls{SED} systems.
Datasets can differ by the temporal resolution of the labels e.g. temporally \emph{strong} or \emph{weak} labels or by the fact that annotator confidence may be present (e.g. \emph{soft} labels) or not, or again, by which sound classes are actually considered during the annotation process. 
To spur research towards addressing these issues, this year task involves two datasets DESED and MAESTRO on which participants systems are benchmarked, while external data and pre-trained models can also be leveraged. 
Due to the aforementioned annotation inconsistencies participants will need to devise novel and effective ways to cope with the fact that sound events that are considered in DESED may be present in MAESTRO but are not annotated and vice versa. 
To ease the challenge participation entry barrier, an updated baseline system was developed. 
Results from such baseline suggest that leveraging more data, if the aforementioned problems are addressed in a reasonable way, is always beneficial. In fact, we show that it is possible to obtain a  system trained on multiple datasets which is stronger than single systems that are trained on each dataset/scenario independently.

% Either list references using the bibliography style file IEEEtran.bst
%\clearpage
\balance
\bibliographystyle{IEEEtran}
\bibliography{refs}

%
% or list them by yourself
% \begin{thebibliography}{9}
% 
% \bibitem{dcase2016web}
%   \url{http://www.cs.tut.fi/sgn/arg/dcase2016/}.
%
% \bibitem{IEEEPDFSpec}
%   {PDF} specification for {IEEE} {X}plore$^{\textregistered}$,
%   \url{http://www.ieee.org/portal/cms_docs/pubs/confstandards/pdfs/IEEE-PDF-SpecV401.pdf}.
%
% \bibitem{PDFOpenSourceTools}
%   Creating high resolution {PDF} files for book production with 
%   open source tools, 
%   \url{http://www.grassbook.org/neteler/highres_pdf.html}.
%
% \bibitem{eWilliams1999}
% E. Williams, \emph{Fourier Acoustics: Sound Radiation and Nearfield Acoustic
%   Holography}. London, UK: Academic Press, 1999.
% 
% \bibitem{ieeecopyright}
%   \url{http://www.ieee.org/web/publications/rights/copyrightmain.html}.
%
% \bibitem{cJones2003}
% C. Jones, A. Smith, and E. Roberts, ``A sample paper in conference
%   proceedings,'' in \emph{Proc. IEEE ICASSP}, vol. II, 2003, pp. 803--806.
% 
% \bibitem{aSmith2000}
% A. Smith, C. Jones, and E. Roberts, ``A sample paper in journals,'' 
%   \emph{IEEE Trans. Signal Process.}, vol. 62, pp. 291--294, Jan. 2000.
% 
% \end{thebibliography}

\end{sloppy}
\end{document}